\documentclass[aps,pra]{revtex4}
\usepackage{graphicx}

\begin{document}

\title{Secure Bidirectional Quantum Communication Protocol without Quantum Channel\\
\thanks{*Email: zhangzj@wipm.ac.cn }}

\author{Z. J. Zhang and Z. X. Man \\
{\normalsize Wuhan Institute of Physics and Mathematics, Chinese
Academy of Sciences, Wuhan 430071, China } \\
{\normalsize *Email: zhangzj@wipm.ac.cn }}

\date{\today}
\maketitle

\begin{minipage}{380pt}
In this letter we propose a theoretical deterministic secure
direct bidirectional quantum communication protocol by using
swapping quantum entanglement and local unitary operations, in
which the quantum channel for photon transmission can be
discarded, hence any attack with or without eavesdropping or even
the destructive attack without scruple is impossible.\\

PACS Number(s): 03.67.Hk, 03.65.Ud\\
\end{minipage}

Much attention [1-6] has been focused on the study of the quantum
key distribution (QKD) after the pioneering work of Bennett and
Brassard published in 1984 [7], for the shared QKD can be used to
encrypt the secret messages which is sent through a classical
channel. As a matter of fact, the deterministic secure direct
communication is more attractive and usually desired due to its
obvious convenience. However, because of its more demanding on the
security than QKDs, a proposal of a deterministic secure direct
communication protocol is usually quite difficult. Till the end of
2003, only three deterministic deterministic secure direct
communication protocols had been proposed by using the quantum
entanglement of a photon pair [4-6]. Moreover, recently Zhang et
al has proposed another deterministic secure direct communication
protocol by using the quantum entanglement swapping of two photon
pairs [8]. These four protocols mentioned are all
message-unilaterally-transmitted protocols. Very recently,
inspired by the deterministic secure direct protocol (i.e., the
ping-pong protocol) proposed by Bostr\"{o}m and Felbinger [5], a
deterministic secure direct bidirectional simultaneous
communication protocol is proposed by Zhang et al in a subtle way
[9]. This is the first bidirectional secure quantum communication
protocol. After this, according to the subtle idea presented in
[9] the improvement on the two-step secure communication protocol
[6] is also finished [10]. Hence, to our best knowledge, there are
only six deterministic secure direct communication protocols so
far. These protocols (except for the one in [8]) have four common
properties as follows. (a) In all the protocols, after the message
sender's encoding by unitary operation on the photon, the photon
must be transmitted form the message sender's side to the message
receiver's side via a quantum channel. This offers opportunities
for the hostile person to eavesdrop or to attack the secret
messages.  (b) All the protocols are essentially quasisecure.
Alternatively, the eavesdropping can not be detected with a
possibility of $100\%$. As a result, a part of information might
be leaked to the eavesdropper. (c) All the protocols are insecure
under the attack without eavesdropping [11]. Hence, a strategy
like message authentification should be adopted to detect such
attacks. (d) In the cases that an eavesdropper is detected or an
attacker attacks the quantum channel of photon transmission
without scruple, the secure communication has to be aborted. In
fact, such cases occur quite possibly during some special times
like a war. We think, all these are the common faults of the
protocols, especially (d) is fatal. By the way, since a quantum
channel for photon transmission should also be included in the
protocol in [8], (d) is also the protocol's fault. In this letter,
taking advantage of the idea of using the quantum entanglement
swapping of two photon pairs [8] and the subtle idea presented in
[9] to realize the bidirectional communication, we further propose
another deterministic secure direct bidirectional simultaneous
communication protocol, however, in which the quantum channel of
photon transmission can be discarded, any attack with or without
eavesdropping and even the destructive attack without scruple are
accordingly impossible, hence this protocol is perfectly secure
and can work in any case provided that the resource of the message
carriers is sufficient. Consequently, this protocol will be very
attractive in the commercial and military aspects due to its
outstanding advantages.

Let us first describe the quantum entanglement swapping [12-14]
simply. Let $|0\rangle$ and $|1\rangle$ be the horizontal and
vertical polarization states of a photon, respectively. Then the
four Bell states, $|\Psi ^{\pm }\rangle =(|01\rangle \pm
|10\rangle )/\sqrt{2}$ and $|\Phi ^{\pm }\rangle =(|00\rangle \pm
|11\rangle )/\sqrt{2}$, are maximally entangled states in the
two-photon Hilbert space.  Let the initial state of two photon
pairs (i.e., the photon $a_1$ and $b_1$ pair and the photon $a_2$
and $b_2$ pair) be the product of any two of the four Bell states,
such as $|\Psi _{a_1b_1}^{+}\rangle$ and $|\Psi
_{a_2b_2}^{+}\rangle$, then after the Bell state measurements on
the photon $a_1$ and $a_2$ pair and the photon $b_1$ and $b_2$
pair, since the following equation holds,
\begin{eqnarray}
|\Psi _{a_1b_1}^{+}\rangle\otimes |\Psi
_{a_2b_2}^{+}\rangle=\frac{1}{2}(|\Psi _{a_1a_2}^{+}\rangle |\Psi
_{b_1b_2}^{+}\rangle-|\Psi _{a_1a_2}^{-}\rangle|\Psi
_{b_1b_2}^{-}\rangle+|\Phi _{a_1a_2}^{+}\rangle|\Phi
_{b_1b_2}^{+}\rangle-|\Phi _{a_1a_2}^{-}\rangle|\Phi
_{b_1b_2}^{-}\rangle),
\end{eqnarray}
the total initial state (i.e., $|\Psi _{a_1b_1}^{+}\rangle \otimes
|\Psi _{a_2b_2}^{+}\rangle$) is projected onto $|\eta _{1}\rangle
=|\Phi _{a_1a_2}^{+}\rangle \otimes |\Phi _{b_1b_2}^{+}\rangle ,
|\eta _{2}\rangle = |\Phi _{a_1a_2}^{-}\rangle \otimes |\Phi
_{b_1b_2}^{-}\rangle , |\eta _{3}\rangle =|\Psi
_{a_1a_2}^{+}\rangle \otimes |\Psi _{24}^{+}\rangle$ and $|\eta
_{4}\rangle =|\Psi _{a_1a_2}^-\rangle \otimes |\Psi _{b_1b_2}^-
\rangle$ with equal probability of $\frac{1}{4}$ for each. It is
seen that previous entanglements between photons $a_1$ and $b_1$,
and $a_2$ and $b_2$, are now swapped into the entanglements
between photons $a_1$ and $a_2$, and $b_1$ and $b_2$.  Therefore,
if $|\Phi _{a_1a_2}^{+}\rangle$ is obtained by the Bell state
measurements, $|\Phi _{b_1b_2}^{+}\rangle$ should be gained
inevitably by the Bell state measurements; if $|\Phi
_{a_1a_2}^-\rangle$ is obtained, then $|\Phi _{b_1b_2}^-\rangle$
is arrived at; and so on. This means that for a known initial
state the Bell state measurement results after the quantum
entanglement swapping are correlated. In the above example $|\Psi
_{a_1b_1}^{+}\rangle \otimes |\Psi _{a_2b_2}^{+}\rangle$ is chosen
as the initial state. In fact, similar results can also be arrived
at provided that other choices of the initial states are given. As
can be seen as follows:
\begin{eqnarray}
\left\{ \begin{array}{c}|\Psi _{a_1b_1}^{+}\rangle\otimes |\Psi
_{a_2b_2}^{-}\rangle=\frac{1}{2}(|\Psi _{a_1a_2}^{+}\rangle |\Psi
_{b_1b_2}^{-}\rangle-|\Psi _{a_1a_2}^{-}\rangle|\Psi
_{b_1b_2}^{+}\rangle-|\Phi
_{a_1a_2}^{+}\rangle|\Phi _{b_1b_2}^{-}\rangle+|\Phi _{a_1a_2}^{-}\rangle|\Phi _{b_1b_2}^{+}\rangle), \\
 |\Psi _{a_1b_1}^{+}\rangle\otimes |\Phi _{a_2b_2}^{+}\rangle=\frac{1}{2}(|\Psi
_{a_1a_2}^{+}\rangle|\Phi _{b_1b_2}^{+}\rangle-|\Psi
_{a_1a_2}^{-}\rangle|\Phi _{b_1b_2}^{-}\rangle+|\Phi
_{a_1a_2}^{+}\rangle|\Psi _{b_1b_2}^{+}\rangle-|\Phi _{a_1a_2}^{-}\rangle|\Psi _{b_1b_2}^{-}\rangle), \\
 |\Psi _{a_1b_1}^{+}\rangle\otimes |\Phi _{a_2b_2}^{-}\rangle=\frac{1}{2}(|\Psi
_{a_1a_2}^{+}\rangle|\Phi _{b_1b_2}^{-}\rangle-|\Psi
_{a_1a_2}^{-}\rangle|\Phi _{b_1b_2}^{+}\rangle-|\Phi
_{a_1a_2}^{+}\rangle|\Psi _{b_1b_2}^{-}\rangle+|\Phi
_{a_1a_2}^{-}\rangle|\Psi _{b_1b_2}^{+}\rangle).
\end{array}\right .
\end{eqnarray}
By the way, for the above four known initial states the
correlation of the Bell state measurement results after the
quantum entanglement swapping is very useful. On the other hand,
it should also be noted that different results by the Bell state
measurements correspond to different initial states for the above
four known initial states. For examples, when $|\Psi
_{a_1a_2}^{+}\rangle$ and $|\Psi _{b_1b_2}^{-}\rangle$ are
obtained by the Bell state measurements, the initial state should
be $|\Psi _{a_1b_1}^{+}\rangle\otimes |\Psi _{a_2b_2}^{-}\rangle$;
when $|\Phi _{a_1a_2}^{+}\rangle$ and $|\Psi _{b_1b_2}^{-}\rangle$
are obtained by the Bell state measurements, the initial state
should be $|\Psi _{a_1b_1}^{+}\rangle\otimes |\Phi
_{a_2b_2}^{-}\rangle$; and so on. Incidentally, this property is
used in our communication protocol. In addition, it is easily
verified that, the four Bell states can be transformed into each
other by some unitary operations, which can be performed locally
with nonlocal effects. For examples: Let $u_0=|0\rangle\langle
0|+|1\rangle\langle1|$, $u_1=|1\rangle\langle1|-|0\rangle\langle
0|$, $u_2=|0\rangle\langle 1|+|1\rangle\langle0|$,
$u_3=|0\rangle\langle1|-|1\rangle\langle0|$, then $|\Psi
_{a_2b_2}^{+}\rangle$ will be in turn transformed into $|\Psi
_{a_2b_2}^{+}\rangle$, $|\Psi _{a_2b_2}^{-}\rangle$, $|\Phi
_{a_2b_2}^{+}\rangle$, $|\Phi _{a_2b_2}^{-}\rangle$ after the
unitary operations $u_0,u_1,u_2,u_3$ on anyone photon (say, the
$b_2$ photon) of the pair, respectively, that is, $u_0 |\Psi
_{a_2b_2}^{+}\rangle=|\Psi _{a_2b_2}^{+}\rangle$, $u_1 |\Psi
_{a_2b_2}^{+}\rangle=|\Psi _{a_2b_2}^{-}\rangle$, $u_2 |\Psi
_{a_2b_2}^{+}\rangle=|\Phi _{a_2b_2}^{+}\rangle$ and $u_3 |\Psi
_{a_2b_2}^{+}\rangle=|\Phi _{a_2b_2}^{-}\rangle$. Assume that each
of the above four unitary operations corresponds to a two-bit
encoding respectively, i.e., $u _{0}$ to '00', $u _{1}$ to '01',
$u _{2}$ to '10' and $u _{3}$ to '11'. Then, taking advantage of
the quantum entanglement swapping and the assumption of the
two-bit codings, a deterministic secure direct bidirectional
communication protocol can be proposed. We show it later.

Alice prepares an ordered $N$ EPR photon pairs in state $|\Psi
\rangle _{ab}=|\Psi ^{+}\rangle =(|0\rangle _{a}|1\rangle
_{b}+|1\rangle _{a}|0\rangle _{b})/\sqrt{2}$ for each and divides
them into two partner-photon sequences $[a_1, a_2, \dots, a_N]$
and $[b_1, b_2, \dots, b_N]$, where $a_i$ $(b_i)$ stands for the
$a$ ($b$) photon in the $i$th photon pair. Then Bob securely takes
the $b$ photon sequence away as a storage for the future use. By
the way, as for how Bob can securely take the $b$ photon sequence
away as a storage, in principle, it is possible in theory.  Maybe
Bob can use some materials [15-16] to store the photons and take
it away just as a baggage. Maybe the photons can be transmitted
through a fiber to Bob's storage during the peaceful and secure
times. In addition, how to maintain the entanglement properties of
the photon pair is also a question. However, all these are only
technological or other theoretical problems [17-18] and beyond our
present theoretical scope. Whenever Alice and Bob want to
communicate secretly with each other (if they want, they can
publicly announce), they can do as follows. Both Alice and Bob
perform the unitary operations on the photons with even (or odd)
orders in their hands according to their secret messages. For
examples, when Alice wants to let Bob securely know her bit string
'011110\dots', according to her bit string she performs $u_1$,
$u_3$, $u_2$ on the $a_2$, $a_4$ and $a_6$ photons respectively,
and so on. When Bob wants to let Alice securely know his bit
string '101100\dots', she performs $u_2$, $u_3$, $u_0$ on the
$b_2$, $b_4$ and $b_6$ photons respectively, and so on. After
their unitary operations they perform their Bell state
measurements and publicly announces their results. That is, Alice
performs in turn her Bell state measurements on the photon $a_1$
and $a_2$ pair, the photon $a_3$ and $a_4$ pair, etc, and publicly
announce the measurement results in order; Similarly, Bob performs
in turn his Bell state measurements on the photon $b_1$ and $b_2$
pair, the photon $b_3$ and $b_4$ pair, etc, and publicly announces
the measurement results in order. Since Alice (Bob) knows which
unitary operations she (he) has performed, then according to Bob's
(Alice's) measurement results publicly announced and her (his)
measurement results, she (he) can conclude Bob's (Alice's) unitary
operations and accordingly extract Bob's (Alice's) bit string (See
Table 1). So far a deterministic direct bidirectional
communication has been proposed. By the way, if one party has no
secret message to communicate, she or he can have two choices. One
is that she or he only performs the unitary operations randomly to
prevent from Eve's eavesdropping. The other choice is that, she (
or he) publicly tells the partner that she (or he) has no secret
message, then she (or he) does not perform any unitary operations
and does not publicly announce the measurement results anymore. In
any case of the above two choices, the communication protocol is
reduced to a deterministic secure direct
message-unilaterally-transmitted communication protocol.

Since Bob has taken securely the $b$ photon sequence away and the
$a$ photon sequence is in Alice's hand, Eve (the eavesdropper) has
no access to any photons, hence she can neither attack nor
eavesdrop the secret messages. Therefore, the present
deterministic direct bidirectional communication protocol is
secure. In all protocols mentioned in this letter, the message
carrier is photon. In all the other protocols except for [8], the
photon encoded (i.e., performed a local unitary operation) has to
be transmitted to the receiver to be performed a local Bell state
measurement of the photon pair, hence the message transmission is
essentially local. While in the present protocol, since the
quantum entanglement swapping of two photon pairs is employed, the
photons encoded need not to be transmitted via a quantum channel
anymore, hence the message transmission is essentially nonlocal.
It is this nonlocality which leads to the nonattackable property
of the present protocol. This is the essential difference between
the present protocol and other protocols. Incidentally, though in
the present protocol the photon is chosen as the message carrier,
the idea of the present protocol is a general one, hence it should
also be suitable for other message carriers. Recently, the
experimental achievement of Bose-Einstein condensates has
attracted many attentions. It is reported that the entangles
states of pairs of atoms can be created and the coherent
transportation of the neutral atoms can be achieved in the optical
lattices [19-20]. Maybe an alternative experimental demonstration
of the present protocol by using the quantum entanglement swapping
of the entangled atom pairs is feasible in the near future.

In summary, we have proposed a theoretical deterministic secure
direct bidirectional quantum communication protocol by using
swapping quantum entanglement and local unitary operations, in
which the quantum channel for photon transmission can be
discarded, hence any attack is impossible and accordingly the
present protocol is perfectly secure.

This work is supported by the National Natural Science Foundation
of China under Grant No. 10304022.

\begin{minipage}{360pt}
\begin{center}
\vskip 1cm Table 1.  Corresponding relations among Bob's and
Alice's Bell state measurement results, the corresponding states
before measurements and Bob's and Alice's unitary $u$ operations
(i.e., the encoding bits).
\\
\begin{tabular}{cccc}  \hline
 $\left\{ \Phi_{a_1a_2}^{+},\Phi _{b_1b_2}^{+}\right\} $ &
 $\left\{ \Psi _{a_1a_2}^{-},\Psi
_{b_1b_2}^{+}\right\} $ & $\left\{ \Psi _{a_1a_2}^{+},\Phi
_{b_1b_2}^{+}\right\} $ & $\left\{ \Psi _{a_1a_2}^{+},\Phi _{b_1b_2}^{-}\right\} $ \\
$\left\{ \Psi _{a_1a_2}^{-},\Psi _{b_1b_2}^{-}\right\} $ &
$\left\{ \Phi _{a_1a_2}^{+},\Phi _{b_1b_2}^{-}\right\} $ &$\left\{
\Psi
_{a_1a_2}^{-},\Phi _{b_1b_2}^{-}\right\} $ & $\left\{ \Psi _{a_1a_2}^{-},\Phi _{b_1b_2}^{+}\right\} $ \\
$\left\{ \Psi _{a_1a_2}^{+},\Psi _{b_1b_2}^{+}\right\} $ &
$\left\{ \Phi _{a_1a_2}^{-},\Phi _{b_1b_2}^{+}\right\} $ &
$\left\{ \Phi
_{a_1a_2}^{+},\Psi _{b_1b_2}^{+}\right\} $ & $\left\{ \Phi _{a_1a_2}^{+},\Psi _{b_1b_2}^{-}\right\} $ \\
$\left\{ \Phi _{a_1a_2}^{-},\Phi _{b_1b_2}^{-}\right\} $ &
$\left\{ \Psi _{a_1a_2}^{+},\Psi _{b_1b_2}^{-}\right\} $ &$\left\{
\Phi _{a_1a_2}^{-}, \Psi _{b_1b_2}^{-}\right\} $ & $\left\{ \Phi
_{a_1a_2}^{-},\Psi _{b_1b_2}^{+}\right\} $ \\ \hline
$\Psi_{a_1b_1}^{+}\otimes \Psi_{a_2b_2}^{+}$&
$\Psi_{a_1b_1}^{+}\otimes \Psi_{a_2b_2}^{-}$ &
$\Psi_{a_1b_1}^{+}\otimes \Phi_{a_2b_2}^{+}$
&$\Psi_{a_1b_1}^{+}\otimes \Phi_{a_2b_2}^{-}$
\\ \hline
\{$u_0^A(00),u_0^B(00)$\} & \{$u_1^A (01),u_0^B(00)$\} & \{$u _2^A
(10), u_0^B(00)$\} & \{$u_3^A(11),u_0^B(00)$\}
\\\{$u_1^A(00),u_1^B(00)$\} & \{$u_0^A (00),u_1^B(01)$\} & \{$u _0^A
(00), u_2^B(10)$\} & \{$u_0^A(00),u_3^B(11)$\}
\\\{$u_2^A(00),u_2^B(00)$\} & \{$u_2^A (10),u_3^B(11)$\} & \{$u _1^A
(01), u_3^B(11)$\} & \{$u_1^A(01),u_2^B(10)$\}
\\\{$u_3^A(00),u_3^B(00)$\} & \{$u_3^A (11),u_2^B(10)$\} & \{$u _3^A
(11), u_1^B(01)$\} & \{$u_2^A(10),u_1^B(01)$\}
\\ \hline
\end{tabular} \\
\end{center}
\end{minipage}
\vskip 1cm

\noindent[1] N. Gisin, G. Ribordy, W. Tittel, and H. Zbinden, Rev.
Mod. Phys. {\bf 74},145 (2002).

\noindent[2] A. K. Ekert, Phys. Rev. Lett. {\bf 67}, 661 (1991).

\noindent[3] D. Bru$\beta$, Phys. Rev. Lett. {\bf 81}, 3018
(1998).

\noindent[4] A. Beige, B. G. Englert, C. Kurtsiefer, and H.
Weinfurter, Acta Phys. Pol. A {\bf 101}, 357 (2002).

\noindent[5] K. Bostrom and T. Felbinger, Phys. Rev. Lett.
{\bf89}, 187902 (2002).

\noindent[6] F. G. Deng, G. L. Long, and X. S. Liu, Phys. Rev. A
{\bf 68}, 042317 (2003).

\noindent[7] C. H. Bennett and G. Brassard, in {\it Proceedings of
the IEEE International Conference on Computers, Systems and Signal
Processings, Bangalore, India} (IEEE, New York, 1984), p175.

\noindent[8] Z. J. Zhang et al, qaunt-ph/0403218.

\noindent[9] Z. J. Zhang et al, qaunt-ph/0403186.

\noindent[10] Z. J. Zhang et al, qaunt-ph/0403215.

\noindent[11] Q. Y. Cai, Phys. Rev. Lett. {\bf 91}, 109801 (2003).

\noindent[12] J. W. Pan, M. Daniell, S. Gasparoni, G. Weihs, and
A. Zeilinger, Phys. Rev. Lett. {\bf86}, 4435 (2001).

\noindent[13] M. Zukowski, A. Zeilinger, M. A. Horne, and  A. K.
Ekert, Phys. Rev. Lett. {\bf71}, 4287 (1993).

\noindent[14] S. Bose, V. Vedral, and P. L. Knight, Phys. Rev. A
{\bf57}, 822(1998).

\noindent[15] D. F. Phillips, A. Flieischhauer, A. Maier, A. L.
Walsworth, and M. D. Lukin, Phys. Rev. Lett. {\bf86}, 783 (2001).

\noindent[16] C. Liu, Z. Dutton, C. H. Behroozi, and L. V. Hau,
Nature (London) {\bf 409}, 490 (2001).

\noindent[17] H. J. Briegel, W. Dur, J. I. Cirac, and P. Zoller,
Phys. Rev. Lett. {\bf81}, 5932 (1998).

\noindent[18] Z. Zhao, T. Yang, Y. A. Chen, A. N. Zhang, and J. W.
Pan, Phys. Rev. Lett. {\bf 90}, 207901 (2003).

\noindent[19] L. M. Duan, A. Sorensen, J. I. Cirac, and P. Zoller,
Phys. Rev. Lett. {\bf85}, 3991 (2000).

\noindent[20] O. Mandel, M. Greiner, A. Widera, T. Rom, T. W.
Hansch, and I. Bloch, Phys. Rev. Lett. {\bf 91}, 010407 (2003).

\end{document}